\documentclass[aps,pra,reprint,superscriptaddress,twocolumn]{revtex4-1}
\usepackage{amssymb}
\usepackage{amsmath}
\usepackage{graphicx}
\usepackage{epstopdf}
\usepackage{times}
\begin{document}
\title{A two-dimensional architecture for fast large-scale trapped-ion quantum computing}
\author{Y.-K. Wu}
\affiliation{Center for Quantum Information, IIIS, Tsinghua University, Beijing 100084, P. R. China}
\author{L.-M. Duan}\email{Corresponding author: lmduan@tsinghua.edu.cn}
\affiliation{Center for Quantum Information, IIIS, Tsinghua University, Beijing 100084, P. R. China}
\date{\today}

\begin{abstract}
Building blocks of quantum computers have been demonstrated in small to intermediate-scale systems. As one of the leading platforms, the trapped ion system has attracted wide attention. A significant challenge in this system is to combine fast high-fidelity gates with scalability and convenience in ion trap fabrication. Here we propose an architecture for large-scale quantum computing with a two-dimensional array of atomic ions trapped at such large distance which is convenient for ion-trap fabrication but usually believed to be unsuitable for quantum computing as the conventional gates would be too slow. Using gate operations far outside of the Lamb-Dicke region, we show that fast and robust entangling gates can be realized in any large ion arrays. The gate operations are intrinsically parallel and robust to thermal noise, which, together with their high speed and scalability of the proposed architecture, makes this approach an attractive one for large-scale quantum computing.
\end{abstract}

\maketitle

Quantum computing has attracted wide research interest over the past few decades because of its potential computational power beyond any classical computers and in particular the exponential speedup for certain tasks \cite{nielsen2000quantum}. Pioneering works have been conducted for the physical implementation of quantum computing in various systems. Two leading platforms, ion trap and superconducting circuit, have demonstrated high-fidelity qubit initialization, readout, single-qubit and two-qubit gates in small to intermediate-scale systems \cite{wineland2003single.ion,PhysRevLett.113.220501,Ballance2016,Gaebler2016,Monz2016,wright2019benchmarking,
Devoret2013superconducting,gambetta2017superconducting,Wendin_2017,google2019quantum.supremacy}. Despite these progress, we are still far from building a quantum computer to solve practical problems, which requires thousands of logical qubits, or millions of physical qubits via current quantum error correction schemes \cite{PhysRevA.86.032324}.

The difficulty in maintaining and operating qubits turns out to increase significantly with the size of the system. State-of-the-art trapped-ion quantum computers mainly focus on the linear structure of ions, which is estimated to be limited to about a hundred ions in a single chain \cite{wineland1998experimental,PhysRevLett.77.3240,clark2001proceedings}. To further scale up the system, several schemes are under active research \cite{Monroe2013}. One is the ion shuttling technique, where ions are guided by adjustable external electric fields to distribute entanglement between different chains \cite{wineland1998experimental,Kielpinski2002,cirac2000scalable.ion}. Another is the photonic quantum network method, where photons are used to generate entanglement between distant ion chains \cite{10.5555/2011617.2011618,RevModPhys.82.1209,PhysRevA.89.022317}. Since these schemes introduce new components to the ion trap quantum computing, they are experimentally more challenging and subjected to speed limit due to the slow quantum wiring step.

Another possible direction is to explore higher dimensional ion systems, which can lead to remarkable increase in the number of qubits. Furthermore, two-dimensional (2D) or three-dimensional (3D) architecture is more convenient for implementation of fault-tolerant quantum error correction. However, in a single RF trap, high-dimensional ion crystals are generally subjected to micromotion. Although theoretical proposals exist for quantum computing in 2D and 3D systems under micromotion \cite{PhysRevA.90.022332,Wang2015,wu2019thesis}, they are much more complicated and experimentally demanding. To bypass this difficulty, an array of microtraps can be used where individual ions or chains of ions are confined in local trapping potentials \cite{cirac2000scalable.ion,2013surface.array}. Due to the limitation of electrode fabrication techniques, distance between different microtraps is generally larger than the ion separation in a single trap. Since entangling gates between ions rely on the spin-dependent dipole-dipole interaction away from their equilibrium positions, which falls inverse cubically with the distance, the gate speed becomes much slower. Fast entangling gates based on the spin-dependent kicks can operate outside of the Lamb-Dicke limit \cite{cirac2003gzc.gate,PhysRevLett.93.100502}, which provides a way to scale up the system \cite{PhysRevLett.93.100502,PhysRevLett.120.220501}. However, the requirement of simultaneous application of tens of strong laser pulses are very challenging experimentally \cite{PhysRevLett.120.220501,J.J.Hope2019fast.gate}.

In this work, we propose an architecture for scalable ion trap quantum computing, where ions form a 2D array with large distance on the order of tens to hundreds of micrometers, for which the trapping potential can be conventionally realized with the current surface trap fabrication technique. Fast entangling gates between adjacent ions can be implemented on the order of microseconds using the available periodic laser pulses \cite{PhysRevLett.93.100502,PhysRevLett.119.230501}.  The gate is robust against thermal fluctuation and we assume only the Doppler cooling for a large ion crystal. The proposed gate is independent of the system size, and can be conveniently implemented by accumulating a fixed number of periodic pulses on the two targeted ions under an appropriate transverse trap frequency. Furthermore, distant entangling gates can be applied in parallel with vanishing crosstalk errors. With these desired features, our proposed architecture provides an attractive approach to large-scale quantum computing.

\begin{figure}[htbp]
  \centering
  \includegraphics[width=0.9\linewidth]{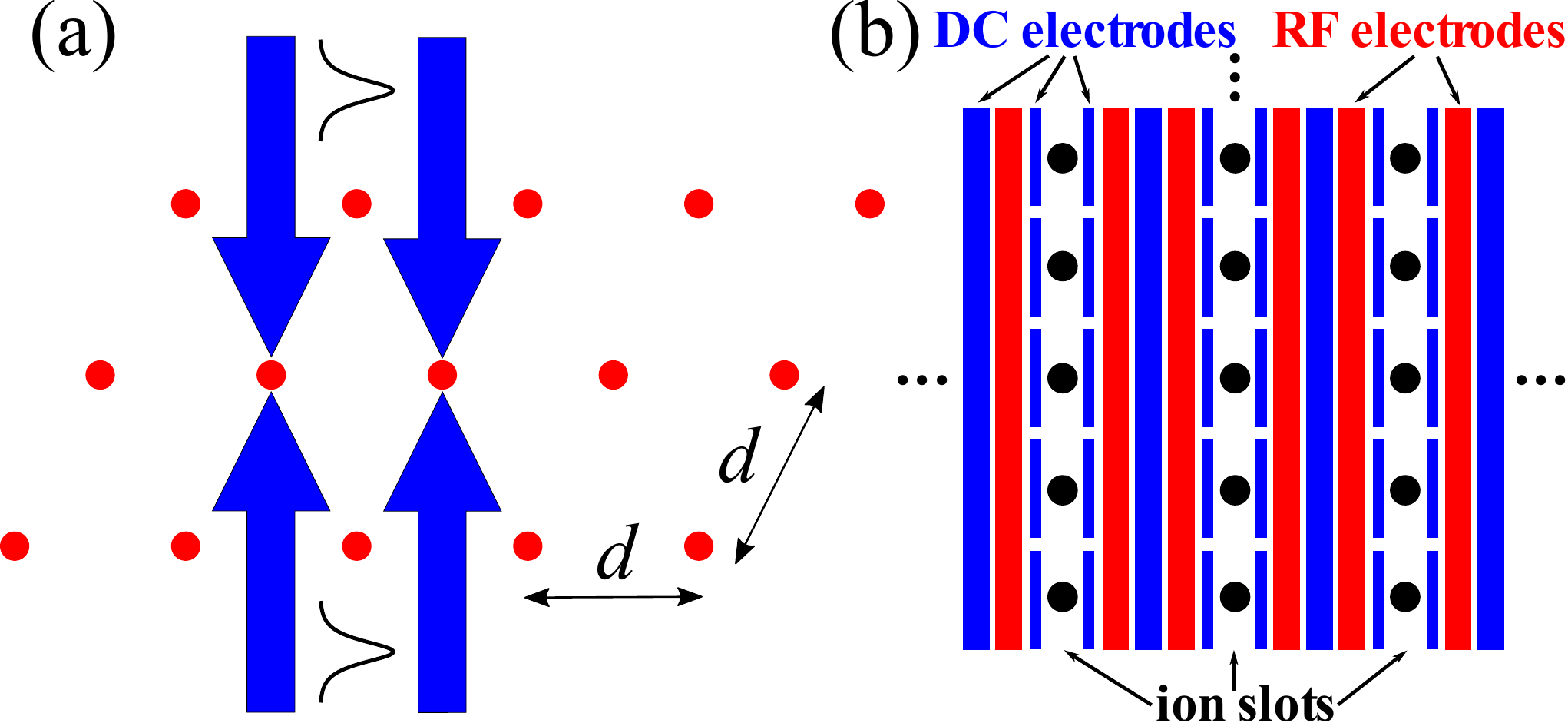}\\
  \caption{Schematic diagram for the proposed ion trap quantum computing architecture. (a) Here we assume that the ions form a square lattice with separation $d$, although generalization to other lattices is straightforward. Each ion feels a harmonic trapping in the $z$ direction perpendicular to the plane with trapping frequency $\omega_z$. Counter-propagating pulsed laser beams are applied on two adjacent ions to entangle them together. (b) The 2D structure of the ions can be realized using e.g. the surface trap technique by first trapping ions into linear chains, and then combing multiple chains together. Ions (black dots in ion slots) and electrodes (blue for DC and red for RF) are not to scale. Due to the size of the electrodes, the ion separation needs to be tens to hundreds of micrometers.}\label{fig:architecture}
\end{figure}

Our architecture is shown schematically in Fig.~\ref{fig:architecture}. Trapped ions form a 2D array with large separation $d$ on the order of tens to hundreds of micrometers. This can be conveniently realized by the available surface trap techniques \cite{surface_trap,2013surface.array} or other microfabrication methods \cite{ouyang2007quadrupole}, e.g., by trapping ions in multiple linear chains aligned in parallel (Fig.~\ref{fig:architecture}(b)), or by trapping each ion in an individual microtrap. In this work we consider square lattice for simplicity, with many of the results can directly be generalized to other types of lattices. Harmonic trapping potential perpendicular to the ion plane is applied on each ion with a trapping frequency $\omega_z$. We require the dipole-dipole interaction between the ions $e^2/4\pi\epsilon_0\sqrt{d^2+(z_1-z_2)^2} - e^2/4\pi\epsilon_0 d \approx -e^2 (z_1-z_2)^2/8\pi\epsilon_0 d^3$ to be much weaker than the transverse trapping energy $m\omega_z^2(z_1^2+z_2^2)/2$, i.e. $\epsilon\equiv e^2/4\pi\epsilon_0 m \omega_z^2 d^3\ll 1$. Due to the 2D nature of the architecture, we only need to consider two-qubit entangling gates on nearest neighbor ions (as well as single-qubit gates on each ion, which can be easily realized) for fault-tolerant universal quantum computing (see e.g. Ref.~\cite{PhysRevA.86.032324}).

To efficiently entangle two ions, we want large spin-dependent displacements of the two ions outside of the Lamb-Dicke regime. Therefore we utilize the spin-dependent kicks (SDKs) \cite{PhysRevLett.110.203001,PhysRevLett.119.230501} generated by two counter-propagating pulsed laser beams along the $z$ direction, as shown in Fig.~\ref{fig:architecture}(a). A fast $\pi$-pulse from the Raman beams can drive the following spin-dependent kick $e^{i\Delta k \hat{z}}\hat{\sigma}_+ + e^{-i\Delta k \hat{z}}\hat{\sigma}_-$ on each ion \cite{cirac2003gzc.gate,PhysRevLett.110.203001,PhysRevLett.119.230501}, where a momentum kick of $\Delta k$ ($-\Delta k$) in the $z$ direction is
accompanied by the internal state change of the ion from $|0\rangle$ to $|1\rangle$ ($|1\rangle$ to $|0\rangle$) and $\Delta k$ denotes the wave vector difference between the two counter-propagating laser beams bridging the Raman transition. The direction of the momentum kicks can be reversed by exchanging the role of two Raman beams (such as their polarizations) \cite{PhysRevLett.119.230501}, which allows us to keep applying spin-dependent momentum kicks in the same direction with the internal state alternating between $|0\rangle$ and $|1\rangle$.

\begin{figure}[htbp]
  \centering
  \includegraphics[width=0.9\linewidth]{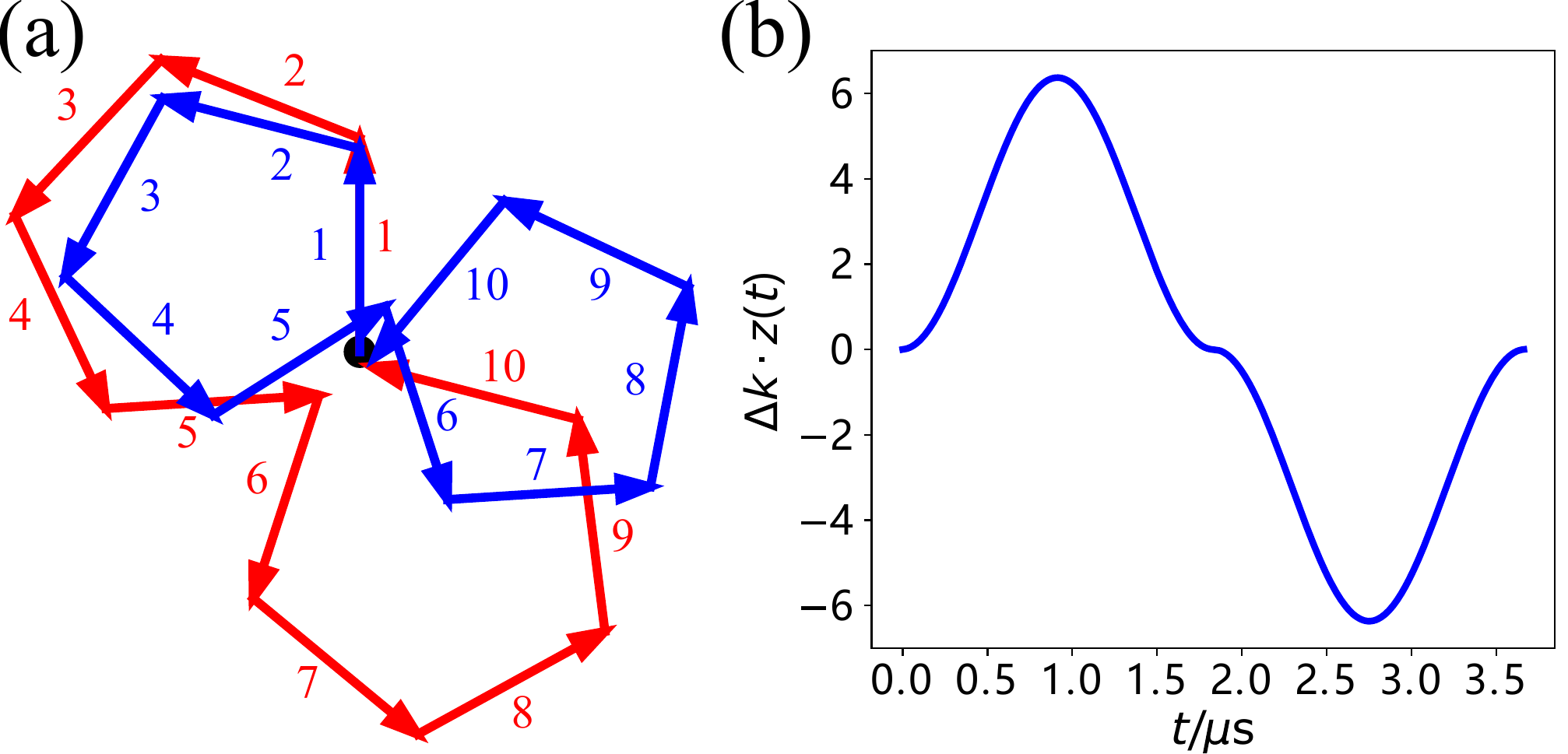}\\
  \caption{(a) Schematic phase-space trajectories for the center-of-mass mode (blue) and the relative mode (red) under the sequence of spin-dependent kicks (SDKs) in an entangling gate. For the two ions initially in $|00\rangle$ or $|11\rangle$, the SDK reads $e^{\pm i\Delta k (\hat{z}_1+\hat{z}_2)}=e^{\pm i\sqrt{2}\eta_c(\hat{a}_c+\hat{a}_c^\dag)} =\hat{D}_c(\pm i\sqrt{2}\eta_c)$, where $\hat{D}_c$ is the displacement operator on the mode $c$ and $\eta_c\equiv\Delta k\sqrt{\hbar/2m\omega_c}$; while for the initial states $|01\rangle$ or $|10\rangle$, the SDK reads $e^{\pm i\Delta k (\hat{z}_1-\hat{z}_2)}=e^{\pm i\sqrt{2}\eta_r(\hat{a}_r+\hat{a}_r^\dag)} =\hat{D}_r(\pm i\sqrt{2}\eta_r)$. Moving into an interaction picture with $H_0=\omega_c\hat{a}_c^\dag \hat{a}_c + \omega_r\hat{a}_r^\dag\hat{a}_r$, these displacement operators further rotate at the frequencies $\omega_c$ and $\omega_r$ respectively, thus the angle between two adjacent displacement operations is $2\pi\omega_{c(r)}/\omega_{\mathrm{rep}}$. Due to the difference between $\omega_c$ and $\omega_r$ (and thus $\eta_c$ and $\eta_r$ as well), the phase-space trajectories do not close after $M=5$ kicks and the endpoints locate on the two sides of the starting point (black dot). By applying a second set of $M$ kicks in the opposite direction, the two trajectories reunite again with the residual displacement suppressed to higher order. (b) $\Delta k\cdot z(t)$ for an ion driven by the SDK sequence $(+M,\,-M)$. The example we consider here is $d=50\,\mu$m for ${}^{171}\mathrm{Yb}^+$ ions with $\omega_z=2\pi\times 0.5444\,$MHz and $M=147$. The trajectory is far outside of the Lamb-Dicke regime, which explains the high gate speed.}\label{fig:gate}
\end{figure}

With the basic tools of SDKs being introduced, here we describe a simple scheme to entangle two nearest-neighbor ions $i$ and $j$ together efficiently. For pedagogical reasons, first we present the gate design for a two-ion crystal by ignoring all the other ions; later we will show that the same gate design works well for the multi-ion case. We will use gate infidelity from the ideal gate $\exp[- i (\pi/4)\hat{\sigma}_z^i \hat{\sigma}_z^j]$ to characterize the gate performance (see Supplementary Materials for details)
\begin{equation}
\delta F = \left(\Theta_{ij} - \frac{\pi}{4}\right)^2 + \sum_k \left(|\alpha_i^k|^2 + |\alpha_j^k|^2\right) \coth \frac{\hbar \omega_k}{2 k_B T_D} \label{eq:infidelity}
\end{equation}
where $\Theta_{ij}$ is the actual two-qubit rotation angle and $\alpha_{i(j)}^k$ is the residual displacement of a collective mode $k$ with frequency $\omega_k$ after the gate due to the laser driving on the ion $i$ ($j$). Throughout this work we will assume the Doppler temperature $k_B T_D = \hbar \Gamma / 2$ where $\Gamma$ is the linewidth of the $D1$ transition.

For a two-ion crystal, the collective motions along the $z$ direction can be described analytically by the center-of-mass mode $\omega_c=\omega_z$ and the relative mode $\omega_r=\omega_z\sqrt{1-2\epsilon}\approx (1-\epsilon)\omega_z$. Now we apply SDKs on the two ions simultaneously. As mentioned above, we can adjust the direction of the kicks such that an ion initially in $|0\rangle$ ($|1\rangle$) keeps feeling $+\Delta k$ ($-\Delta k$) kicks during the gate.
For two ions initially in the $|00\rangle$ ($|11\rangle$) or the $|01\rangle$ ($|10\rangle$) states, the center-of-mass mode or the relative mode will be excited. If we keep applying momentum kicks on the two ions at a repetition rate of $\omega_{\mathrm{rep}} \gg \omega_z$, the trajectories of the two modes will be close to two circles in the phase spaces (see Fig.~\ref{fig:gate}(a)) with radii of about $R_{c(r)}=(\sqrt{2}\eta_{c(r)}/2\pi) \omega_{\mathrm{rep}}/\omega_{c(r)}$, where $\eta_{c(r)}\equiv\Delta k \sqrt{\hbar/2m\omega_{c(r)}}$ is the Lamb-Dicke parameter. Since $\omega_c\approx\omega_r$, we expect the two trajectories to close roughly at the same time $T=2\pi/\omega_z$, that is, after about $M=\omega_{\mathrm{rep}}/\omega_z$ pulses.
The accumulated phase for each initial spin state after the displacement operations is proportional to the area enclosed by the phase-space trajectory. For a maximal entangling gate, we want the phase difference between the two modes $\Delta\phi = 2\pi R_r^2 - 2\pi R_c^2
\approx 3\epsilon \hbar \Delta k^2 \omega_{\mathrm{rep}}^2/2\pi m\omega_z^3$, hence between the two sets of states,
to be $2\Theta_{ij}=\pi/2$.
Since the two frequencies $\omega_c$ and $\omega_r$ are not exactly the same, there will be a residual displacement error of about $\pi\epsilon \times (\sqrt{2}\eta_z/2\pi) \omega_{\mathrm{rep}}/\omega_z = \sqrt{\pi\epsilon\Delta\phi/6}=O(\sqrt{\epsilon})$ on each mode. We can further suppress this error by applying SDKs in the pattern of $(+M,\,-M)$, namely first $M$ momentum kicks in one direction followed by $M$ kicks in the opposite directions (see Fig.~\ref{fig:gate}(a)). Then the overall effect on a mode with frequency $\omega$ is given by $\sum_{n=0}^{M-1} e^{i 2\pi n \omega/\omega_{\mathrm{rep}}} -
\sum_{n=M}^{2M-1} e^{i 2\pi n \omega/\omega_{\mathrm{rep}}} = \left(1-e^{i2\pi M\omega/\omega_{\mathrm{rep}}}\right) \sum_{n=0}^{M-1} e^{i 2\pi n \omega/\omega_{\mathrm{rep}}}$, which is reduced by a factor of $1-e^{i2\pi M\omega/\omega_{\mathrm{rep}}} \approx 2\pi i \left(1 - M\omega/\omega_{\mathrm{rep}}\right) = O(\epsilon)$. In this case we need $\Delta\phi=\pi/4$ for a maximal entangling gate, that is
\begin{equation}
\omega_z = \left( \frac{3e^2\hbar\Delta k^2 \omega_{\mathrm{rep}}^2}{2\pi^3\epsilon_0 m^2 d^3} \right)^{\frac{1}{5}}
\label{eq:omz}
\end{equation}
We can extend the pattern to $(+M,\,-2M,\,+M)$ to suppress displacement errors to higher order, but it turns out that for the examples considered in this work, the pattern of $(+M,\,-M)$ suffices, with a gate infidelity of $O(\epsilon^3)$.

In the experiment, $\omega_{\mathrm{rep}}$ is usually fixed or can only be adjusted in a small range; while for our proposed architecture, $d$ is also fixed by the design of the local trapping potential. From Eq.~(\ref{eq:omz}) we see that the required $\omega_z$ is thus determined for the desired entangling gate. However, this can lead to an additional error due to the discrete nature of momentum kicks because $M=\omega_{\mathrm{rep}}/\omega_z$ may not be an integer. To minimize this discretization error, we shift $\omega_z$ around Eq.~(\ref{eq:omz}) slightly such that
\begin{align}
M =& \frac{\omega_{\mathrm{rep}}}{2\omega_z} + \frac{\omega_{\mathrm{rep}}}{2(1-\epsilon)\omega_z} \approx \left(1 + \frac{\epsilon}{2}\right)\frac{\omega_{\mathrm{rep}}}{\omega_z} \nonumber\\
=& \left(1 + \frac{1}{2} \frac{e^2}{4\pi\epsilon_0 d^3 m \omega_z^2}\right)\frac{\omega_{\mathrm{rep}}}{\omega_z}
\label{eq:M}
\end{align}
is an integer. Depending on the initial value of $\omega_z$ we computed from Eq.~(\ref{eq:omz}), we may need to round $M$ by at most $1/2$. Therefore the relative change in $\omega_z$ can be bounded by $1/2M$, which corresponds to a relative change in $\Delta\phi$ of $5\Delta\omega_z/\omega_z\le5/2M$, an absolute under- or over-rotation of $5\pi/4M$ of the gate, or a round-off infidelity upper bounded by $(5\pi/8M)^2$. This coherent error may be alleviated by alternatively rounding $M$ upwards or downwards; also note that if we are able to adjust $\omega_{\mathrm{rep}}$ slightly, or if we design $d$ carefully in the beginning, we can completely remove this error.

As a numerical example, we consider ${}^{171}\mathrm{Yb}^+$ ions driven by counter-propagating $355\,$nm pulsed lasers, with $\omega_{\mathrm{rep}}=2\pi\times 80\,$MHz \cite{PhysRevLett.119.230501}; we take $\Gamma=2\pi\times 20\,$MHz to estimate the Doppler temperature. For $d=50\,\mu$m, we need $M=147$ and $\omega_z=2\pi\times 0.5444\,$MHz. This corresponds to $\epsilon=0.00056$, and a gate time of $T=3.675\,\mu$s. The gate fidelity for the two-ion crystal is $F_2=1-1.5\times 10^{-5}$ using Eq.~(\ref{eq:infidelity}), which contains both the residual displacement and the round-off error. For $d=250\,\mu$m, we similarly get $M=386$, $\omega_z=2\pi\times 0.2073\,$MHz, $\epsilon=3.1\times 10^{-5}$, $T=9.65\,\mu$s and $F_2=1-3.0\times10^{-5}$. Such a gate speed is faster than many current typical trapped ion platforms, with tremendously larger ion separation. This is owing to our gate scheme
with displacements far outside of the Lamb-Dicke regime. In Fig.~\ref{fig:gate}(b) we plot the meanfield trajectory of an ion under the SDK sequence for the $d=50\,\mu$m case. The large value of $\Delta k \cdot z(t)$ clearly indicates the breakdown of the Lamb-Dicke approximation, thus the advantage of the SDK method.

\begin{figure}[htbp]
  \centering
  \includegraphics[width=0.9\linewidth]{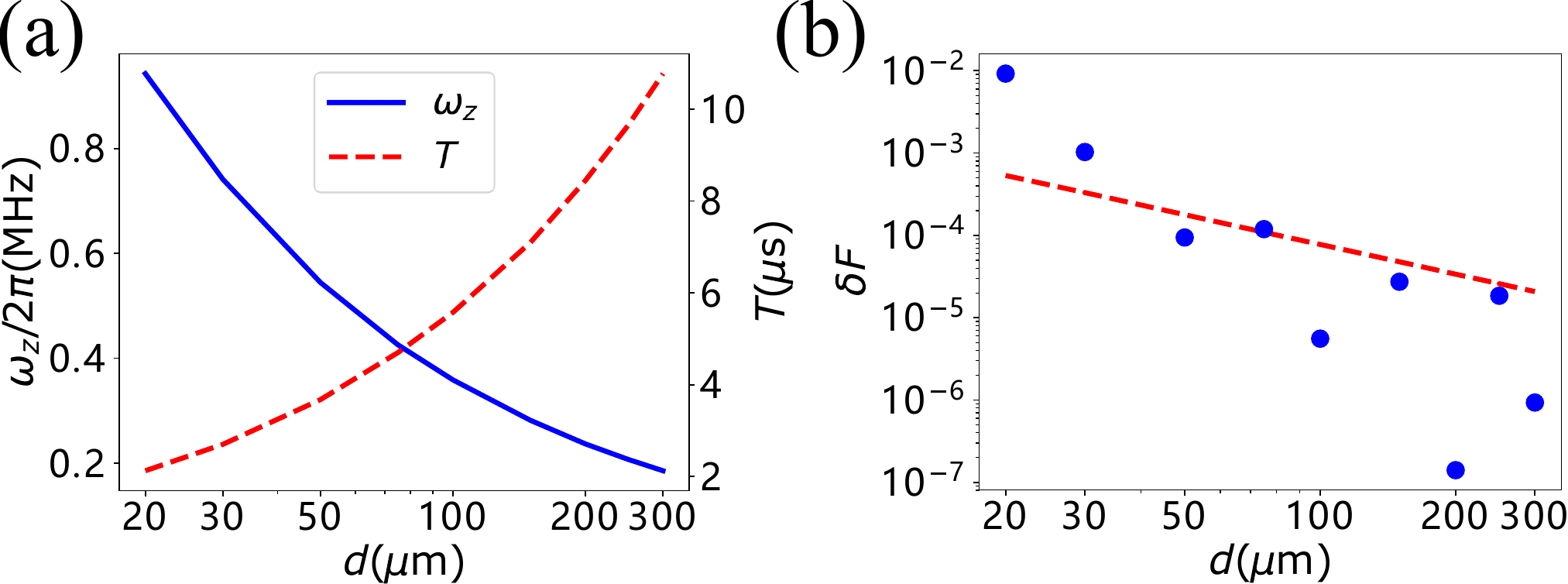}\\
  \caption{(a) Required transverse trapping frequency $\omega_z$ (solid blue line) and the time for maximal entangling gate $T$ (red dashed line) vs. ion separation $d$ in the 2D square lattice. The gate parameters are completely determined by the ion separation and the SDK pattern $(+M,\,-M)$ (note that $M$ varies with $d$). (b) Numerically computed gate infidelity from Eq.~(\ref{eq:infidelity}) for two central ions in a $10\times 10$ array (blue dots) and the theoretical upper bound on the error of rounding $M$ to an integer (red dashed line). At small $d$, the gate infidelity is dominated by the residual displacement of each phonon mode; while for large $d$ the round-off error becomes more important. Note that the round-off error can be much smaller than the theoretical upper bound for carefully chosen $d$.}\label{fig:param}
\end{figure}

Now a major advantage of our proposed architecture and gate scheme is that, the above gate design for the two-ion crystal can directly be applied to the multi-ion 2D array without significant change in performance. For two central ions in a 2D square lattice of $10\times 10$ ions, the same gate design for $d=50\,\mu$m gives a fidelity of $F_{10\times 10}=1-8\times 10^{-5}$, and for $d=250\,\mu$m a fidelity of $F_{10\times 10}=1-2.6\times 10^{-5}$. We further study the scaling of the gate time $T$ and the required transverse trapping frequency $\omega_z$ versus ion separation $d$ in Fig.~\ref{fig:param}(a), with some typical gate infidelity on the $10\times 10$ lattice presented in Fig.~\ref{fig:param}(b). For $d$ in the range of 30 to 250 micrometers, we generally obtain $\omega_z$ of hundreds of kilohertz, $T$ of a few microseconds, and a gate infidelity below $10^{-3}$.

This insensitivity to the system size, i.e. scalability, is owing to the combination of large ion distance and fast gate speed in our architecture, so that the response of far-away ions can be safely neglected. Actually for an infinite 2D square lattice, the collective modes are travelling waves with frequencies distributed in a narrow band of $O(\epsilon \omega_z)$. Hence the propagation speed of local perturbation can be characterized by the group velocity $|\boldsymbol{v}_g| < 3.5\epsilon\omega_z d$ (see Supplementary Materials).
Then it takes $O(1/\epsilon\omega_z)$ time for a local disturbance to propagate to a nearby ion, and for our gate time of $O(1/\omega_z)$, the response of other ions are $O(\epsilon)$ smaller, thus can be neglected. More discussion about this propagation speed can be found in Supplementary Materials. Note that this slow propagation is not conflicted with the fast gate speed: when applying an entangling gate, both targeted ions are driven with large displacement; while for the propagation speed we are considering the response of one ion due to the disturbance on another one.

\begin{figure}[htbp]
  \centering
  \includegraphics[width=0.9\linewidth]{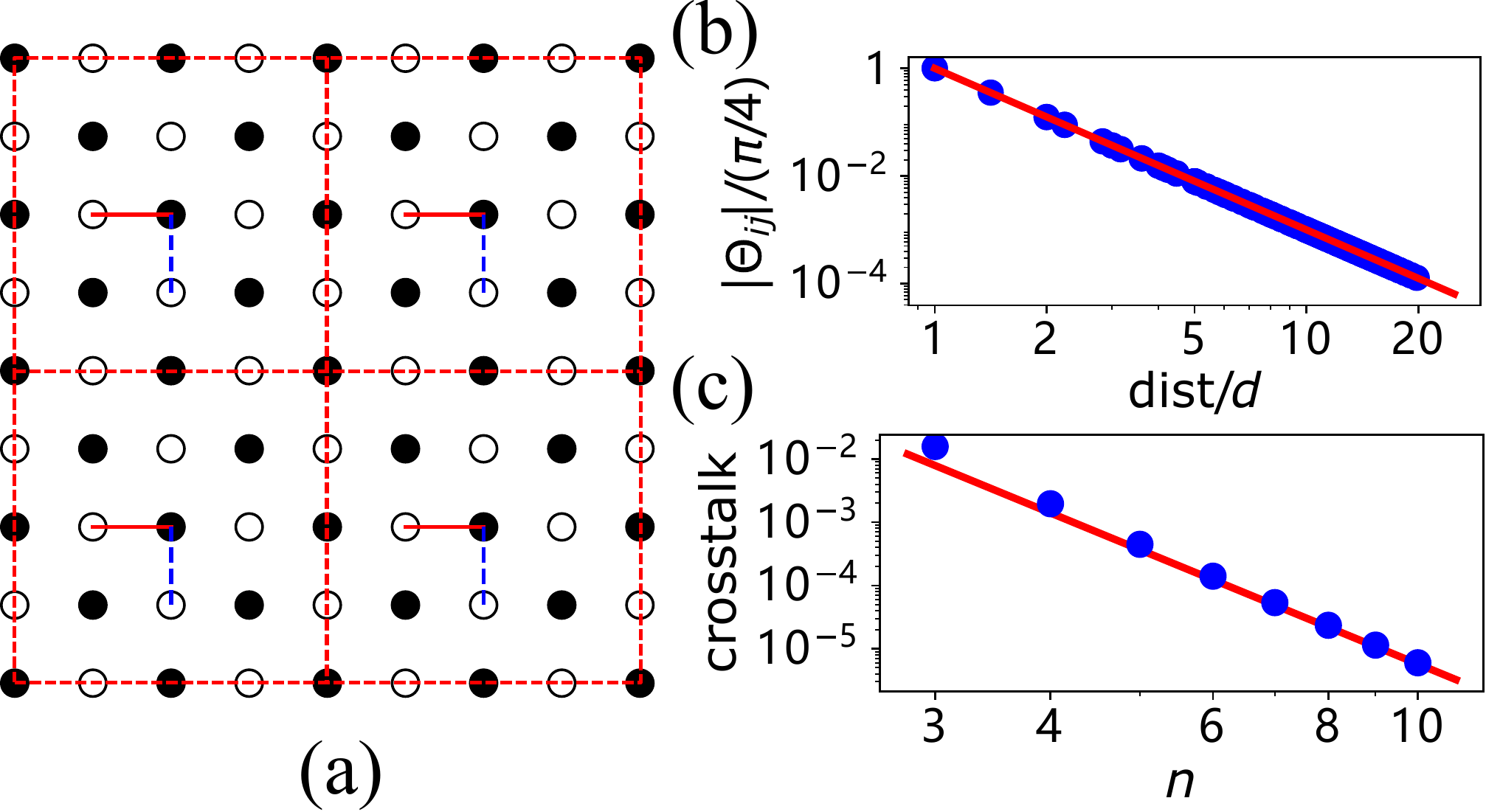}\\
  \caption{(a) Our proposed architecture directly supports a surface-code-like pattern of parallel entangling gates. Here the white and the black dots represent the data and the measurement qubits. We divide the 2D array into $(n+1)\times (n+1)$ blocks (red dashed boxes, for clearance we choose $n=4$ here to demonstrate the idea). Corresponding entangling gates in different blocks (e.g. the nearest-neighbor pairs connected by the red solid lines or by the blue dashed lines) can be performed in parallel; while gates within the same block need to be executed in serial. Therefore there is only a polynomial overhead $O(n)$ compared with the original surface code \cite{PhysRevA.86.032324}. (b) The crosstalk error of applying SDKs on two ions $i$ and $j$ is a two-qubit rotation term $\exp(-i\Theta_{ij}\sigma_z^i\sigma_z^j)$ \cite{wu2019parallel}. The numerically computed $\Theta_{ij}$ on a $31\times31$ array (blue dots) agrees well with the theoretical inverse cubic relation with the distance between the two ions (red line) \cite{wu2019parallel}. Its contribution to gate infidelity can be estimated as $|\Theta_{ij}|^2$. (c) The crosstalk error per gate (blue dots) when parallelizing gates in all the blocks in (a) vs. block size $n$. When $n$ is large, the numerically computed crosstalk error approaches a theoretical value of $5.75/n^6$. For the crosstalk error per gate to be lower than $10^{-3}$, we need $n\ge 5$.}\label{fig:parallel}
\end{figure}

Furthermore, such fast high-fidelity entangling gates can be applied in parallel on distant ion pairs. We have shown in Ref.~\cite{wu2019parallel} how entangling gates can be parallelized in uniform 1D chain or 2D hexagonal lattice, and similar derivations can be made here for the square lattice (see Supplementary Materials for details). With the 2D structure of the proposed architecture, a surface-code-like entangling gate pattern \cite{PhysRevA.86.032324} is natively supported, as shown in Fig.~\ref{fig:parallel}(a). By dividing the 2D array into $(n+1)\times (n+1)$ blocks with shared edges, entangling gates in different blocks can be performed in parallel; on the other hand, entangling gates within the same block needs to be performed in serial, which gives $O(n)$ overhead for all the nearest-neighbor entangling gates. As shown in Ref.~\cite{wu2019parallel}, the crosstalk error of applying laser sequence on two distant ions $i$ and $j$ is a two-qubit rotation $\exp(-i\Theta_{ij}\hat{\sigma}_z^i\hat{\sigma}_z^j)$. In Fig.~\ref{fig:parallel}(b) we numerically compute $\Theta_{ij}$ for two ions at different positions of the lattice and verify the $O(1/r_{ij}^3)$ scaling where $r_{ij}$ is the separation between the two ions (in the unit of $d$). The gate infidelity $\delta F=|\Theta_{ij}|^2$, in a similar form as Eq.~(\ref{eq:infidelity}), is thus $O(1/r_{ij}^6)$. Now we can add up the crosstalk infidelity for all the ion pairs driven simultaneously and study the scaling with $n$, as shown in Fig.~\ref{fig:parallel}(c). When $n$ gets large, the numerically computed value approaches the analytical form of $5.75/n^6$ (see Supplementary Materials). When $n\ge 5$, we get a crosstalk error per gate lower than $10^{-3}$.

We want to point out that this architecture is also robust against small fluctuation in local trapping frequencies.
For a shift of $\delta\omega_z$, we expect an infidelity due to over- or under-rotations of about $(5\pi\delta\omega_z/4\omega_z)^2$.
There is also
a displacement error of about $\sqrt{2}\eta_z M \delta\omega_z/\omega_z$ after $M$ pulses, which is further reduced by about $2\pi \delta\omega_z/\omega_z$ after the first-order error cancellation; thus we estimate an infidelity of about $8\pi^2\eta_z^2 M^2(\delta\omega_z/\omega_z)^4$.

In this paper we use ${}^{171}\mathrm{Yb}^+$ for numerical examples because pulsed laser and SDKs are already utilized in this system \cite{PhysRevLett.110.203001,PhysRevLett.119.230501}. However, the proposed architecture can also work well for other species of ions. In Supplementary Materials we further present some numerical examples for ${}^9\mathrm{Be}^+$ and ${}^{40}\mathrm{Ca}^+$.

In summary, we have proposed a 2D ion trap architecture with large ion spacing for fast and scalable quantum computing. The gate design is universal and parallel for any large ion array with a negligible intrinsic gate infidelity even under the Doppler temperature. The required elements of this architecture have been demonstrated in ion trap experiments \cite{PhysRevLett.119.230501}. Given its scalability and convenience of the associated trap fabrication, the proposed architecture provides a promising route for large-scale quantum computing.

\begin{acknowledgments}
This work was supported by the National key Research and Development Program of China (2016YFA0301902), the Frontier Science Center for Quantum Information of the Ministry of Education of China, and the Tsinghua University Initiative Scientific Research Program. Y.K.W. acknowledges in addition support from Shuimu Tsinghua Scholar Program and the International Postdoctoral Exchange Fellowship Program.
\end{acknowledgments}

%
\end{document}


\title{Supplementary Materials for ``A two-dimensional architecture for fast large-scale trapped-ion quantum computing''}
\author{Y.-K. Wu}
\affiliation{Center for Quantum Information, IIIS, Tsinghua University, Beijing 100084, P. R. China}
\author{L.-M. Duan}\email{Corresponding author: lmduan@tsinghua.edu.cn}
\affiliation{Center for Quantum Information, IIIS, Tsinghua University, Beijing 100084, P. R. China}

\maketitle
\section{Transverse dynamics of a 2D ion crystal}
\label{sec:mode}
Suppose $N$ ions are trapped by some external potential in a 2D array with equilibrium positions $(x_i,\, y_i)$ ($i=1,\,2,\,\cdots,\,N$). Here we consider the transverse motion of the ions in the $z$ direction with a strong harmonic trapping $\omega_z$. If the displacements of the ions $|z_i|$ are much smaller than their minimal distance $d$, the (classical) equation of motion is given by
\begin{equation}
m \ddot{z}_i = -m\omega_z^2 z_i + \frac{1}{2} \frac{e^2}{4\pi\epsilon_0}\sum_{j\ne i} \frac{z_i-z_j}{[(x_i-x_j)^2 + (y_i-y_j)^2]^{3/2}},
\end{equation}
from which we can solve the transverse dynamics. We can regard it as a matrix equation
\begin{equation}
m \frac{d^2\boldsymbol{z}}{dt^2} = - V \boldsymbol{z} \label{eq:dynamics},
\end{equation}
with the potential matrix given by
\begin{equation}
V_{ij} = \left\{
\begin{array}{cc}
\frac{e^2}{4\pi\epsilon_0}\frac{1}{[(x_i-x_j)^2 + (y_i-y_j)^2]^{3/2}} & (i \ne j)\\
m \omega_z^2 - \frac{e^2}{4\pi\epsilon_0} \sum_{k\ne i}\frac{1}{[(x_i-x_k)^2 + (y_i-y_k)^2]^{3/2}} & (i = j)
\end{array}
\right..
\end{equation}
We can diagonalize this matrix to solve the collective modes of the ions and their time evolution; or we can quantize these collective modes for designing entangling gates.

In particular, for a square lattice with separation $d$, we define its lattice vectors
\begin{equation}
\boldsymbol{a}_1 = d\left(1,\,0,\,0\right), \qquad \boldsymbol{a}_2 = d\left(0,\,1,\,0\right),
\end{equation}
with the corresponding reciprocal vectors
\begin{equation}
\boldsymbol{b}_1 = \left(1,\,0,\,0\right), \qquad \boldsymbol{b}_2 = \left(0,\,1,\,0\right).
\end{equation}

The position of an ion on the 2D lattice can now be represented by two integer indices $\alpha$ and $\beta$ as $\boldsymbol{r}_{\alpha\beta} = \alpha \boldsymbol{a}_1 +\beta \boldsymbol{a}_2$.
Plugging this expression into the potential matrix, we get
\begin{equation}
V_{\alpha\beta,\alpha'\beta'} = \left\{
\begin{array}{cc}
\frac{e^2}{4\pi\epsilon_0 d^3} \frac{1}{[(\alpha-\alpha')^2 + (\beta-\beta')^2]^{3/2}} & (\alpha \ne \alpha' \textrm{ or } \beta\ne \beta')\\
m \omega_z^2 - \frac{e^2}{4\pi\epsilon_0 d^3} \sum_{(\mu,\lambda)\ne(\alpha,\beta)}\frac{1}{[(\alpha-\mu)^2 + (\beta-\lambda)^2]^{3/2}} & (\alpha=\alpha',\, \beta=\beta')
\end{array}
\right..
\end{equation}

For a finite number of ions, we use the above method to find the collective modes. In the main text we also consider the limit of infinite number of ions. In this case the collective modes in the $z$ direction are travelling waves described by the wave vector $\boldsymbol{k}=k_1 \boldsymbol{b}_1 + k_2 \boldsymbol{b}_2$ ($k_1, k_2 \in (-\pi/d,\pi/d]$), with a mode vector
\begin{equation}
z_{\alpha\beta}^{\boldsymbol{k}}(t) \propto \exp \left[ i(\boldsymbol{k}\cdot \boldsymbol{r}_{\alpha\beta}-\omega_{\boldsymbol{k}} t) \right] = \exp \left[ i(\alpha k_1 d + \beta k_2 d -\omega_{\boldsymbol{k}} t) \right].
\end{equation}
The mode frequency can be solved by substituting this mode vector into the equation of motion. We get
\begin{align}
\omega_{\boldsymbol{k}}
=& \omega_z \sqrt{1 - \epsilon {\sum_{\alpha,\beta}}' \frac{1-\cos (\alpha k_1 d + \beta k_2 d)}{(\alpha^2+\beta^2)^{3/2}}}\\
\approx& \omega_z \left[1 - \epsilon {\sum_{\alpha,\beta}}' \frac{1-\cos (\alpha k_1 d + \beta k_2 d)}{2(\alpha^2+\beta^2)^{3/2}}\right],
\end{align}
where the notation $\sum_{\alpha,\beta}'$ means a summation over all integer pairs of $\alpha$ and $\beta$ apart from when they are both zero.
\section{Numerical computation of gate fidelity}
\label{sec:fidelity}
Suppose we want to entangle the ions $i$ and $j$ within the $N$-ion crystal by applying $M$ simultaneous spin-dependent momentum kicks on them with a fixed repetition rate. For the initial state of an ion in $|0\rangle$, suppose the direction of the kicks are $s_1,\,s_2,\,\cdots,\,s_M$ ($s_i=\pm 1$) along the $z$ direction (and for an ion initially in $|1\rangle$ the directions are $-s_1,\,-s_2,\,\cdots,\,-s_M$). Because the evolution of the internal states is completely determined as they are flipped by each SDK, we can focus on the evolution of the motional states. Then the computation of the gate fidelity is very similar to that of a Molmer-Sorensen-like gate under continuous-wave (CW) laser driving (see e.g. Ref.~\cite{wu2018noise.analysis}). All what we need is to replace the CW case $\Omega(t)\sin (\mu t+\varphi)$ by a series of delta functions $\sum_{l=1}^M s_l \delta(t-t_l)$ where $t_l$ is periodic and denotes the arrival time of the $l$-th SDK. Following the notation of Ref.~\cite{wu2018noise.analysis}, we have
\begin{equation}
\alpha_j^k = i \eta_k b_j^k \sum_{l=1}^M s_l e^{i\omega_k t_l} \label{eq:alpha},
\end{equation}
and
\begin{equation}
\Theta_{ij} = -2\sum_k \eta_k^2 b_i^k b_j^k \sum_{l=2}^M \sum_{m=1}^{l-1} s_l s_m \sin \omega_k(t_l-t_m) \label{eq:theta}.
\end{equation}
Note that here we have an additional negative sign in $\Theta_{ij}$ compared with the definition in Ref.~\cite{wu2018noise.analysis} to make it positive. The unitary time evolution is
\begin{equation}
U = \exp\left[-i\Theta_{ij}\hat{\sigma}_z^i\hat{\sigma}_z^j + \sum_k(\alpha_i^k\hat{\sigma}_z^i + \alpha_j^k\hat{\sigma}_z^j)a_k^\dag - (\alpha_i^{k*}\hat{\sigma}_z^i + \alpha_j^{k*}\hat{\sigma}_z^j)a_k \right].
\end{equation}

The gate infidelity for an initial state $|+\rangle|+\rangle$ is then
\begin{equation}
\delta F = \left(\Theta_{ij} - \frac{\pi}{4}\right)^2 + \sum_k \left(|\alpha_i^k|^2 + |\alpha_j^k|^2\right) \coth \frac{\hbar \omega_k}{2 k_B T},
\end{equation}
where $T$ is the temperature of the initial thermal distribution of the motional state. We can verify that, when restricted to the two-ion case, this result is consistent with Ref.~\cite{PhysRevLett.119.230501} after averaging over the thermal state. As we show in Ref.~\cite{wu2018noise.analysis}, the average gate infidelity over all initial states is related to this ``worst case'' infidelity by a factor of $4/5$.

\section{Parallel entangling gates}
\label{sec:parallel}
In Ref.~\cite{wu2019parallel} we have studied the crosstalk error on a 1D uniform chain and a 2D hexagonal lattice. The crosstalk error of addressing two ions $i$ and $j$ simultaneously is a two-qubit rotation term $\exp(-i\Theta_{ij}\hat{\sigma}_z^i\hat{\sigma}_z^j)$, with $\Theta_{ij}$ decaying inverse cubically with the distance between the two ions.

The derivation for 2D square lattice is very similar.
The difference from the hexagonal lattice is that, now for the square lattice, the norm of a displacement vector $\boldsymbol{r}_{\alpha\beta} = \alpha \boldsymbol{a}_1 +\beta \boldsymbol{a}_2$ is $d \sqrt{\alpha^2+\beta^2}$ instead of $d \sqrt{\alpha^2+\beta^2+\alpha\beta}$.
With this substitution, the rest of the derivation is the same as Appendix~A of Ref.~\cite{wu2019parallel}.
Finally we get a scaling of $1/(n^2+m^2)^{3/2}$ for the two-qubit rotation angle, or $1/(n^2+m^2)^{3}$ for the crosstalk gate infidelity when parallelizing two entangling gates separated by the displacement vector $\boldsymbol{r}_{nm}$.

What we get above is the crosstalk error for two gates, while in Fig.~4 of the main text we are trying to parallelizing all the entangling gates on a sublattice with separation $nd$. The infidelity of these two-qubit rotation terms on different ion pairs simply add up together. For each pair of entangling gates to be parallelized, we have four crosstalk terms among the four involved ions. When $n$ is large, we can ignore the small difference in the distance of these four terms (for example, when the four ions are on the same line, the actual distances are $n-1$, $n$, $n$ and $n+1$). Then we need to evaluate
\begin{equation}
{\sum_{\alpha,\beta}}' \frac{1}{n^6} \frac{1}{(\alpha^2+\beta^2)^{3}} \approx \frac{4.659}{n^6}.
\end{equation}
Note that every two entangling gates have four ion pairs for the crosstalk error, and that such crosstalk errors are shared by these two gates. Also note that at $n=1$ the ``crosstalk error'' is just the entangling gate we want to realize with $\Theta_{ij}=\pi/4$. We finally get the crosstalk error per entangling gate as
\begin{equation}
\left(\frac{\pi}{4}\right)^2 \times 4 \times \frac{1}{2} \times \frac{4.659}{n^6} \approx \frac{5.75}{n^6},
\end{equation}
which is valid for large $n$.
\section{Propagation of local disturbance}
\label{sec:propagate}
\begin{figure}[htbp]
  \centering
  \includegraphics[width=0.5\linewidth]{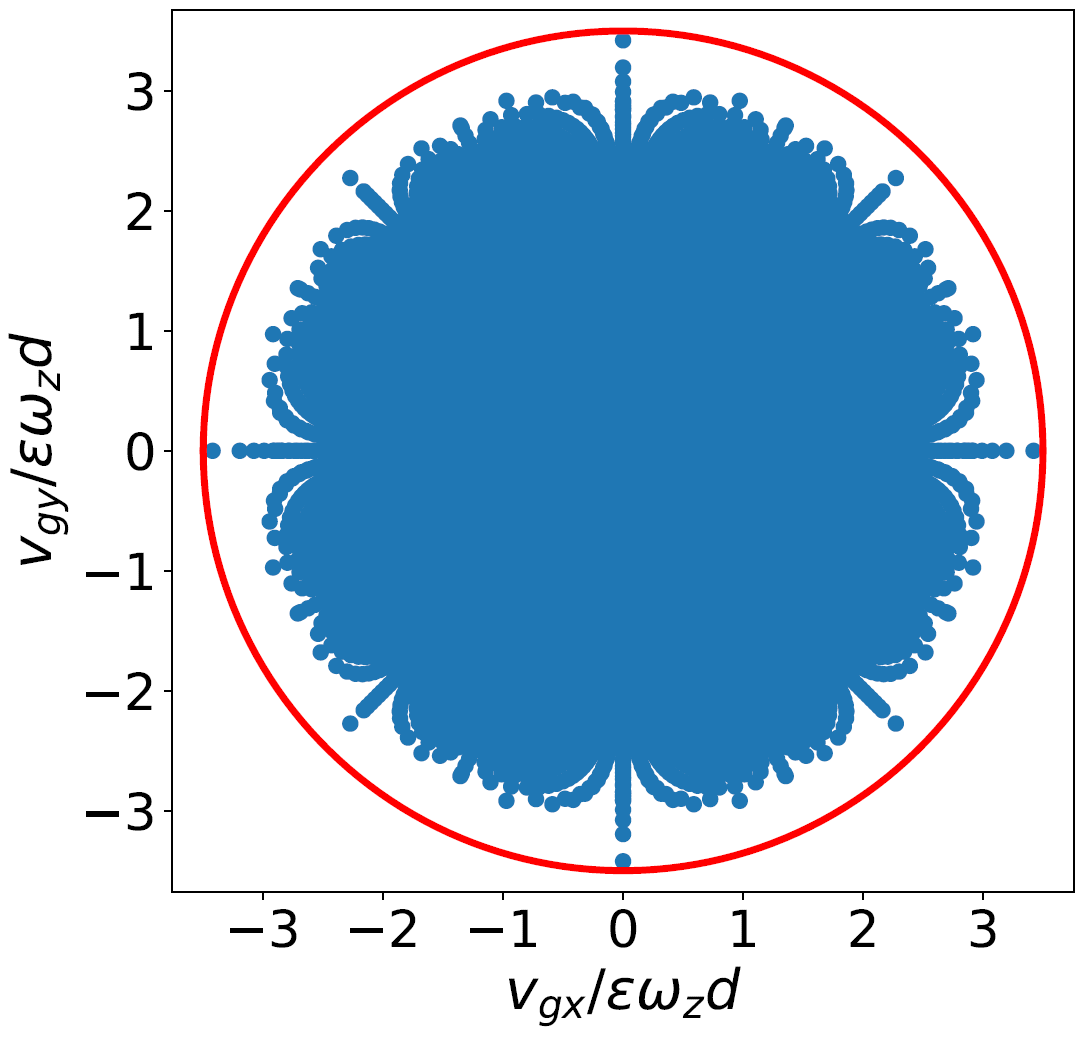}\\
  \caption{Group velocity in a $201\times201$ square lattice with $\epsilon\ll 1$. The red circle indicates a maximal group velocity of about $v_g\approx 3.5\epsilon \omega_z d$.}\label{fig:velocity}
\end{figure}

Consider an infinite square lattice. According to Sec.~\ref{sec:mode}, the transverse modes are travelling waves with frequencies distributed in a narrow band of $O(\epsilon\omega_z)$. We can thus use the group velocity to characterize the speed of propagation
\begin{equation}
\boldsymbol{v}_g(\boldsymbol{k}) = \nabla_{\boldsymbol{k}} \omega(\boldsymbol{k})
\approx -\frac{\epsilon\omega_z d}{2} {\sum_{\alpha,\beta}}' \frac{\sin (\alpha k_1 d + \beta k_2 d)}{(\alpha^2+\beta^2)^{3/2}} \left(\alpha,\,\beta\right), \label{eq:velocity}
\end{equation}
where we use the fact $\epsilon\ll 1$.
In Fig.~\ref{fig:velocity} we numerically evaluate Eq.~(\ref{eq:velocity}) for each $\boldsymbol{k}$ on a $201\times 201$ square lattice as the blue dots. They are bounded by a maximal group velocity of about $3.5\epsilon\omega_z d$.

Strictly speaking, the speed of signal in this system is the speed of light $c$, which we take as infinity when writing down an electrostatic Coulomb potential. Here we further study the response of far-away ions outside the ``light cone'' defined by $v_g$.

Suppose we have an initial disturbance at $t=0$ on the central ion to give it a displacement $z_0$ and a velocity $v_0$, while all the other ions stay at rest. Having solved the collective modes, we can express the response of other ions as
\begin{align}
z_{nm}(t) =& \mathrm{Re} \frac{d^2}{4\pi^2} \int_{-\pi/d}^{\pi/d} dk_1 \int_{-\pi/d}^{\pi/d} dk_2 \left[z_0 + \frac{i v_0}{\omega(\boldsymbol{k})}\right] e^{i[n k_1 d+m k_2 d - \omega(\boldsymbol{k})t]} \nonumber\\
\approx & \mathrm{Re} \frac{d^2}{4\pi^2} \left[z_0 + \frac{i v_0}{\omega_z}\right] \int_{-\pi/d}^{\pi/d} dk_1 \int_{-\pi/d}^{\pi/d} dk_2 e^{i[n k_1 d+m k_2 d - \omega(\boldsymbol{k})t]} \label{eq:response}
\end{align}
where we use the fact that $\omega(\boldsymbol{k})\approx \omega_z$ for $\epsilon\ll 1$ and only keep the $\boldsymbol{k}$ dependence in the phase factor.

Now we study the scaling of the above expression vs. $n$ and $m$. Similar to our calculation for parallel gates and the derivations in Appendix~A of Ref.~\cite{wu2019parallel}, with some algebra, we can decompose the above integral into several terms like
\begin{equation}
\sin \{\omega_z t [1-\epsilon \zeta]\} \int_{-\pi}^{\pi} dx \int_{-\pi}^{\pi} dy \cos[\lambda S(x,y)] \cos (nx + my)
\end{equation}
where $\lambda\equiv \epsilon \omega_z t$, $S(x,y)\equiv \sum_{\alpha,\beta}' \cos (\alpha x + \beta y)/2(\alpha^2+\beta^2)^{3/2}$, $\zeta\equiv S(0,0)$ and $x=k_1 d$, $y=k_2 d$. In this expression the first term corresponds to a fast oscillation at the local trap frequency; while the terms in the integral describe the slow change of the pulse shape as it propagates.

From Ref.~\cite{wu2019parallel} and Sec.~\ref{sec:parallel} we know that this integral scales as $1/(n^2+m^2)^{3/2}$ for large $n$ and $m$. With similar derivations, we can bound the other terms appearing in the original integral in Eq.~(\ref{eq:response}) and finally the amplitude is also bounded by $1/(n^2+m^2)^{3/2}$.
Furthermore, this scaling shall become valid when $n$ or $m$ is much larger than $\lambda=\epsilon \omega_z t$, which is the only other scale defined in the above expression. In other words, this $1/(n^2+m^2)^{3/2}$ scaling is valid on the sites where the ``light cone'' defined by the group velocity has not reached.

Therefore we expect a response of $(v_g t / d) (n^2+m^2)^{-3/2}$ for small $t$. Then the total response of other ions at a gate time of $O(1/\omega_z)$ is $O(\epsilon)$. This explains the good performance of our gate scheme in the multi-ion case.
\section{Further numerical examples}
Here we consider ${}^9\mathrm{Be}^+$ and ${}^{40}\mathrm{Ca}^+$ as two further numerical examples. Since current experiments do not use pulsed laser for these systems, here we assume a laser frequency red-detuned to the $D1$ transition of the ion with a repetition rate $\omega_{\mathrm{rep}}=2\pi\times 80\,$MHz. Also we use $\Gamma=2\pi\times 20\,$MHz to estimate the Doppler temperature. For ${}^9\mathrm{Be}^+$ we consider a laser wavelength of $318\,$nm and get (1) $d=50\,\mu$m, $M=43$, $\omega_x=2\pi\times 1.861\,$MHz, $\epsilon=0.0009$, $T=1.075\,\mu$s, $F_{10\times 10}=0.9994$; and (2) $d=250\,\mu$m, $M=114$, $\omega_x=2\pi\times 0.7018\,$MHz, $\epsilon=5.1\times 10^{-5}$, $T=2.85\,\mu$s, $F_{10\times10}=1-1.9\times 10^{-5}$.
For ${}^{40}\mathrm{Ca}^+$ we use a wavelength of $400\,$nm and obtain (1) $d=50\,\mu$m, $M=86$, $\omega_x=2\pi\times 0.9306\,$MHz, $\epsilon=0.00081$, $T=2.15\,\mu$s, $F_{10\times10}=0.9998$; and (2) $d=250\,\mu$m, $M=227$, $\omega_x=2\pi\times 0.3524\,$MHz, $\epsilon=4.5\times 10^{-5}$, $T=5.675\,\mu$s, $F_{10\times10}=1-4\times 10^{-5}$.

%